# Resonant random laser emission from graphene quantum dot doped dye solutions

Peymaneh Rafieipour[1], Abbas Ghasempour Ardakani[1,*], Fatemeh Daneshmand[2]

[1]Department of Physics, Shiraz University, Shiraz 71454, Iran
[2]Department of Physics, Kharazmi University, Tehran, Iran
*Corresponding author: aghasempour@shirazu.ac.ir

## Abstract

Graphene quantum dots (GQDs) are more promising than other kinds of semiconductor QDs because of their photostability and biocompatibility in different applications such as bioimaging, biosensing and light emitting diodes (LEDs). In addition, advances in random lasers (RLs) have led to an emerging desire for developing remote sensing and detecting strategies, lightning and imaging systems that are far cheaper, more precise and simpler. Although combining GQDs and RLs seems promising for the development of advanced biosensing and bioimaging systems, the RLs fabricated based on GQDs have been rarely studied. Here, we report on the fabrication of dye doped GQDs RLs with resonant feedback that are pumped optically with nanosecond pulses. GQDs, synthesized by the pyrolysis of citric acid, are used as scattering centers in an ethylene glycol solution of rhodamine B dye. It is demonstrated experimentally that discrete lasing modes with subnanometer linewidths appear at pump fluences above the threshold. Furthermore, the dependence of random lasing emission characteristics on the concentration of GQDs and the pump position is investigated experimentally.

## 1. Introduction

Random laser (RL) is a research topic of ongoing relevance and importance, especially with regards to the development of new low cost and mirrorless laser technologies [1]. It has gained much attention due to its potential applications in lightning, bioimaging, sensing and detecting [1]. The observation of an emission light with lasing characteristics has been attracting much interest in the sense that it comprises a highly disordered gain medium. This phenomenon was first observed by lawandy et. al. in colloidal solutions including rhodamine 640 perchlorate dye and $TiO_2$ particles in methanol [2]. It was demonstrated that light multi-scattering inside a random amplifying medium can result in an intensive and narrow-band lasing emission when pump energy approaches a threshold. Later in 1999, a multi-mode random lasing emission was reported in ZnO powder [3]. Cao et. al. provided an explanation of multi-mode RL emission based on recurrent light scattering and the formation of closed loop paths served as optical cavities [3]. These observations indicate that light scattering is the basis of realizing random lasing emission and the strength of light multi-scattering in a disordered active medium does indeed affects many of RL emission characteristics. Considering different feedback mechanisms for light in a disordered amplifying medium, RLs are divided in two categories: RLs with non-resonant feedback and RLs with resonant feedback [4]. In the former category, gain line-shape of the active medium determines the lasing frequency due to the absence of light interference effects and phase information in the diffusive medium. One recent theory suggested that the overlap of many simultaneously oscillating lasing modes results in the singled-peak and relatively broad emission of RLs with non-resonant feedback [5]. On the contrary for the latter case, resonant frequencies of the random cavities determine the mode frequencies of the RL emission spectrum. It is because light localization in a strongly scattering medium preserves the phase information due to constructive interference events of light. These emerging light sources have fascinated researches for the past few decades considering the variety of papers on manufacturing and characterizing different kinds of RLs [6-11].

Recent developments on graphene-based RLs (GRLs) highlight the role of graphene in realizing a new class of RLs suitable for designing high-performance optoelectronic and nanoelectronic devices [12-16]. Graphene which is a two-dimensional honeycomb lattice of $sp^2$ hybridized carbon atoms is well-known by its fascinating properties that can be tuned through chemical doping, external magnetic fields and applied voltages. In a recent study on GRLs, a highly porous vertical-graphene-nanowalls (GNWs) network is employed to multiscatters the emission light from perovskite nanocrystals and provides the essential optical feedback for realizing RL action with an ultra-low threshold energy density [13]. It was shown that under excitation with a 374 nm pulsed laser, sharp peaks of linewidth ~0.4 nm appear at the emission spectra for pump powers above a threshold which demonstrates the occurrence of RL action. Furthermore, coating the GNWs by $Ag/SiO_2$ was demonstrated to lower the threshold pump fluence due to combined effect of high reflectivity and plasmonic enhancement of Ag. In 2017, an ultra-low threshold electrically pumped RL was achieved from a devise composed of graphene quantum dots (GQDs) sandwiched between two graphene layers on top of a $p-Si/SiO_2$ substrate and on a polymethyl-methacrylate (PMMA) layer [14]. The nonlinear change of the emission intensity and linewidth with external bias provides a signature for the occurrence of RL action which was accompanied by the appearance of very narrow peaks with subnanometer linewidths in the emission spectrum. Graphene structures have been also applied in RLs for tunning and enhancing the RL performance. H. Fujiwara et. al. employed graphene flakes to switch and tune a ZnO NP-based-RL with white light illumination [15]. In another similar attempt in 2012, hybrid structure of graphene oxide nanoflakes/ZnO nanorods was suggested to enhance the RL action from ZnO nanorods by the assistance of graphene surface plasmons [16].

There are, in general, a number of types of papers that investigate the use of semiconductor QDs (SQDs) as the gain or scattering medium in RLs [17-21]. However, the preparation method of SQDs consists of rather complex and sophisticated procedures. On the other hand, considering the increasing advances in related bioimaging and biosensing applications, there is a demand for developing low toxic and more photostable QDs. In this way, graphene quantum dots (GQDs) are new and emerging carbon-based QDs which have received considerable attention for their unique properties and widespread applications. They are more photostable, biocompatible and environmentally friendly than conventional SQDs such as CdS, CdSe, CdTe and PbS [22]. Because of their outstanding properties, they can be regarded as a promising replacement of SQDs in many applications including light emitting diodes (LEDs), solar cells, bioimaging, biosensing and photocatalysis [22]. While developing GQD-based RLs seems interesting from a practical point of view, using GQDs in RLs has rarely been investigated. Up to our knowledge, there is only one report on RLs fabricated based on GQDs [23]. In 2013, lasing emission with RL characteristics was reported from a Fabry-Perot cavity containing a mixture of GQDs and $TiO_2$ nanoparticles as the gain and scattering medium, respectively [23]. In that research, GQDs were fabricated by laser ablation method and compared with the same functionalized carbon dots (C-dots). Regardless of the sophisticated, boring and expensive preparation method, it was shown that there is an improved lasing performance and a 5-fold enhancement of optical gain for GQDs over the same functionalized C-dots.

Herein, we demonstrate the fabrication of dye doped GQD RLs with resonant feedback. In our proposed structure, GQDs act as the scattering medium in order to provide optical feedback. Rhodamine B (RhB) dye is also used as the gain medium to provide optical amplification. GQDs are synthesized by a simple and inexpensive method based on direct pyrolysis of citric acid. It is worth mentioning that this method provides us with a much better and robust control over the size and nature of carbon-based products (e.g. GQD or GO) [24]. Other advantageous of this method to be chosen for synthesizing GQDs are the large mass production, fast solubility in water and high stability in air at room temperature. We then characterized the synthesized GQDs by transmission electron microscope (TEM), UV-Vis absorption, Fourier Transform Infrared (FTIR) and Photoluminescence (PL) analysis. It was shown previously that one can use the water solution of this synthesized GQD for diagnosis of exhale acetone in diabetes Mellitus, due to its special structural properties as well as high photostability, low toxicity, bright luminescence and biocompatibility [25]. Very recently, a similar study on acetone detection is reported based on GQDs functionalized three-dimensional ordered mesoporous ZnO [26]. In this letter, we fabricate a dye doped GQD RL in which discrete lasing modes with FWHM of ~1 nm or less appear at pump fluences above the threshold. It is shown that GQDs can be used as the scattering medium in a RL system. In addition, the dependence of the RL emission spectrum on the excitation position and the concentration of GQDs are investigated experimentally. Our results are well-consisted with the theory of RLs with resonant feedback.

**Experimental method**

GQDs are synthesized by the pyrolysis of citric acid [24]. Citric acid was purchased from Merck (monohydrate, ACS, ISO, Reag). In a similar method to [24], we transferred 5.000 g citric acid into a beaker. Then, it was placed on a hot plate and heated to 203°C at air temperature. Citric acid melted at the temperature of 136°C and turned into pale yellow at 161°C. After 45 minutes passed, the beaker contained an orange dense liquid of GQDs. Fig. 1 (a) exhibits a TEM image of the synthesized GQDs. Scale bar is 44.8 nm. One can estimate the average diameter of 5.93 nm for GQDs. UV-Vis and PL spectra of the synthesized GQDs are depicted in Fig. 1 (b). PL spectrum was recorded under a 360 nm excitation source. It is inferred from Fig. 1 (b) that GQDs mostly emit light at 475 nm under UV excitation. Furthermore, the sharp fall in the UV-Vis spectrum suggests the absorption edge of 490 nm for the synthesized GQDs.

The gain medium was a solution of 8.770 mM RhB dye (sigma Aldrich) in ethylene glycol (Merck, 99.5% purity) which was prepared in a separate beaker after 90 minutes stirring on a magnetic stirrer. To prepare the disordered gain medium composed of dye doped GQDs, we added 2 cc solution of the gain medium into the beaker containing GQDs. Then, the solution of RhB dye and GQDs was mixed together under vigorous stirring for about 90 minutes. The prepared solution with the concentration of 11.90 M GQDs was signed as sample 1. For preparing the RL cell, the prepared solution was then transferred into a cuvette with dimensions of $44.30 \times 17.20 \times 0.05$ mm$^3$.

In RL experiments, we utilized an Nd-YAG pulsed laser as the excitation source which was operating at a pulse width of 10 ns, pulse repetition rate of 10 Hz and the wavelength of 532 nm. A sketch of the experimental set up is shown in Fig. 2. The pump light passes through an aperture diaphragm and focused on the RL cell by a cylindrical lens of 15 cm focal length. The dimension of the excitation stripe on the RL sample was $10.30 \times 0.35$ mm$^2$ which was controlled and adjusted by the aperture diaphragm. An optical fiber was then applied for collecting the RL light which was emitted from the sample during the pumping process. Then, the emitted light was guided and coupled to a spectrometer (Ocean Optics, 0.1 nm resolution). Finally, the emitted RL spectrum can be visualized and analyzed by using a computer.

**Results and discussions**

Fig. 3 depicts the emission spectra corresponding to the sample containing 8.770 mM RhB doped 11.90 M GQDs (sample 1) and that of neat dye at a pump fluence of 114.84 mJ/cm$^2$. The broadband emission spectrum of the neat dye exhibits a PL spectrum with FWHM and the center wavelength of 64 and 598 nm, respectively. On the other hand, there appear about five lasing modes with wavelengths of 620.9, 623.4, 625.0, 627.5 and 629.6 nm, respectively in the emission spectrum of sample 1. Their FWHMs are 1.1, 0.7, 1.1, 1.0 and 0.9 nm, respectively.

The significant difference between the emission spectra of sample 1 and that of neat dye in Fig. 3 indicates that the presence of GQDs in dye solution results in the resonant and multi-mode RL emission from sample 1. The appearance of lasing modes imply on the excitation of resonant cavities formed inside the pump region. One can interpret this observation by considering the theory of RLs with resonant feedback [27]. Typically, the pump light multi-scatters from GQDs and finally returns back to a scatterer from which it was scattered before. Thus, a closed path forms if the returned light interferes constructively. Since the pump light confines to this closed path, it may be regarded as a cavity. The condition for constructive interference of light determines the resonant frequencies of the formed random cavities. There are many such cavities with different losses and quality factors that are distributed randomly inside the gain medium. On the other hand, the confined light is amplified by the gain medium via stimulated emission process during pumping. When optical gain exceeds loss for resonant frequencies of a special cavity or a set of cavities, lasing oscillation occurs at the corresponding resonant modes.

The lasing behavior of sample 1 is further confirmed in Fig. 4. The pump fluence varies from 5.83 mJ/cm$^2$ to 223.6 mJ/cm$^2$. The nonlinear variation of the spectrally integrated output intensity versus pump fluence is an important evidence for laser oscillation. It is shown in Fig. 4 (a) that the RL threshold can be estimated as ~70 mJ/cm$^2$. It is noteworthy to mention here that the strong mode competition between the excited resonant modes results in the lowering of the output intensity and is responsible for the observed fluctuations in Fig. 4 (a) at pump fluences above the threshold.

Fig. 4 (b) displays emission spectra of sample 1 at pump fluences before and after threshold. One can observe that the broadband PL emission is dominant for very low pump fluence of 5.83 mJ/cm$^2$. The corresponding FWHM and central wavelength are approximately 56 and 608 nm, respectively. By increasing the pump fluence, the amount of gain increases which results in decreasing the linewidth of the emission spectrum. Furthermore, the light amplification enhances by increasing the dwell time of emission light in the disordered amplifying medium. Hence, at the pump fluence of 47.71 mJ/cm$^2$ before threshold, we observe a relatively narrow component with FWHM and central wavelength of 11.6 and 611.3 nm, respectively on top of the broadband PL spectrum. By further increasing the pump fluence, the penetration depth of the pump light as well as the amount of gain increases. At threshold pump fluence when the increased gain balances with the losses of the random cavities inside the gain medium, lasing oscillation occurs in resonant frequencies of the cavities. As a result, some very narrow lasing modes appear in the emission spectrum. For example the third spectrum in Fig. 4 (b) which corresponds to the pump fluence of 92.09 mJ/cm$^2$ exhibits the simultaneous oscillation of five lasing modes with wavelengths of 620.3, 622.4, 624.7, 626.8 and 629.1 nm. Their corresponding FWHMs are 1.3, 0.5, 1.1, 0.8 and 1.2 nm, respectively. At higher pump fluences, those resonant cavities which possess higher loss are also activated which result in the excitation of more resonant lasing modes and the occurrence of mode competition effect.

In order to investigate the role of randomness in the RL emission from sample 1, we change the excitation position on the sample and measure the corresponding emission spectra. The results for six different pump positions at the constant pump fluence of 114.84 mJ/cm$^2$ are shown in Fig. 5. They are shifted vertically for better comparison. The obtained emission spectra from down to top are attributed to positions 1 to 6, respectively. It turns out that changing the pump position has a strong impact on the RL emission spectrum. One can observe that the number of lasing modes and their wavelengths change when pump light illuminates different positions on the sample. Furthermore, new lasing modes may oscillate and appear in other regions of the emission spectrum, by changing the pump position on the sample. Typically shown in Fig. 5, there are five lasing modes with wavelengths of 620.9, 623.4, 625.0, 627.5 and 629.6 nm in the emission spectrum for the case of position 1. However, the emission spectrum is singled mode at positions 2 and 3. Their wavelengths are 628.3 and 624.7 nm, respectively. Their relatively large FWHMs of 3.6 and 2.9 nm may be attributed to the mode competition effect and the overlap of simultaneously oscillating lasing modes. In positions 3 and 4, the emission spectrum may be regarded to be almost double mode with pairs of wavelengths ~620.9, 626.2 nm and 623.4, 627.0 nm, respectively. Finally in position 6, there excite five lasing modes with wavelengths of 611.1, 614.4, 619.6, 623.7 and 627.5 nm in the sample.

The dependence of the RL emission spectrum on the excitation position is well-explained by the theory of RLs with resonant feedback [27]. Since resonant cavities are distributed randomly inside the gain medium, different cavities are excited at each pumping process which illuminates a specific position on the sample. Considering the different quality factors and resonant modes of cavities, the emission spectrum differs when different positions on the sample are pumped.

To verify whether the scattering of light from GQDs provides the optical feedback for random lasing emission, we change the concentration of GQDs and investigate the associated RL emission characteristics. Fig. 6 shows the RL emission spectra of samples containing 11.90, 16.65 and 21.41 M GQDs (samples 1, 2 and 3, respectively) at the constant pump fluence of 237.17 mJ/cm$^2$. It should be noted that the excitation position corresponding to the emission spectra of Fig. 6 is different with those corresponding to the emission spectra of Fig. 5. It is inferred from Fig. 6 (a) that the output intensity and the number of lasing modes increases by increasing the concentration of GQDs. There is one lasing mode with wavelength of 625.7 nm in the emission spectrum of sample 1. While for the case of sample 2, three lasing modes with wavelengths of 621.6, 625.2 and 628.5 nm oscillate in the emission spectrum. For sample 3, there appear four lasing modes with wavelengths of 612.1, 616.2, 618.8 and 623.9 nm in the emission spectrum. It is because multiple light scattering enhances by increasing the concentration of scattering elements. As a result, the probability of light trapping inside the gain medium enhances which imply on the reduction of cavity losses. In fact, those cavities which have been inactive before can also excite in highly concentrated solutions of dye doped GQDS. Hence, the number of resonant random ring cavities and consequently the lasing modes in the RL emission spectrum increases.

It is then expected that the RL threshold decreases by increasing the concentration of GQDs (the scattering elements). Fig. 7 (a) depicts and compares the spectrally integrated output intensity versus pump fluence for samples 1, 2 and 3. It is clearly observed that the RL threshold decreases by increasing the concentration of GQDs. The RL thresholds are approximated as ~70, 56 and 39 mJ/cm$^2$ for samples 1, 2 and 3, respectively. Moreover, emission spectra of samples 3, 2 and 1 before and after respective thresholds are also presented in figures 7 (b) to (d), respectively. For sample 3, there appear two narrow lasing modes with wavelengths of 613.9 and 619.6 nm at the pump fluence of 47.71 mJ/cm$^2$. Their corresponding FWHMs are 1 and 0.7 nm, respectively. In the case of sample 2, a well-distinguishable mode with wavelength of 620.3 nm survives in the mode competition and oscillates at the pump fluence of 92.09 mJ/cm$^2$. Its FWHM is 1.8 nm. For sample 1, two lasing modes with wavelengths of 620.9 and 626.2 nm oscillates at the pump fluence of 131.48 mJ/cm$^2$. Their FWHMs are 1.8 and 2.3 nm, respectively. Thus, it is confirmed that the scattering of light from GQDs provides the essential optical feedback for random lasing emission.

**Conclusion**

In summary, we fabricated dye doped GQD RLs with resonant feedback. It was demonstrated experimentally that GQDs can be used as the scattering medium in a RL system. We observed the appearance of discrete lasing modes with FWHMs of ~1 nm or less in the emission spectrum, for pump fluences above the threshold. Furthermore, the dependence of RL emission characteristics on the excitation position and concentration of GQDs was confirmed experimentally. It was shown that the scattering of light from GQDs provided the necessary optical feedback for random lasing emission.

**Acknowledgment**

We thank Dr. S. Salmani, Dr. M. H. Majlesara and Saeed BehAeen.

**Figure Captions**

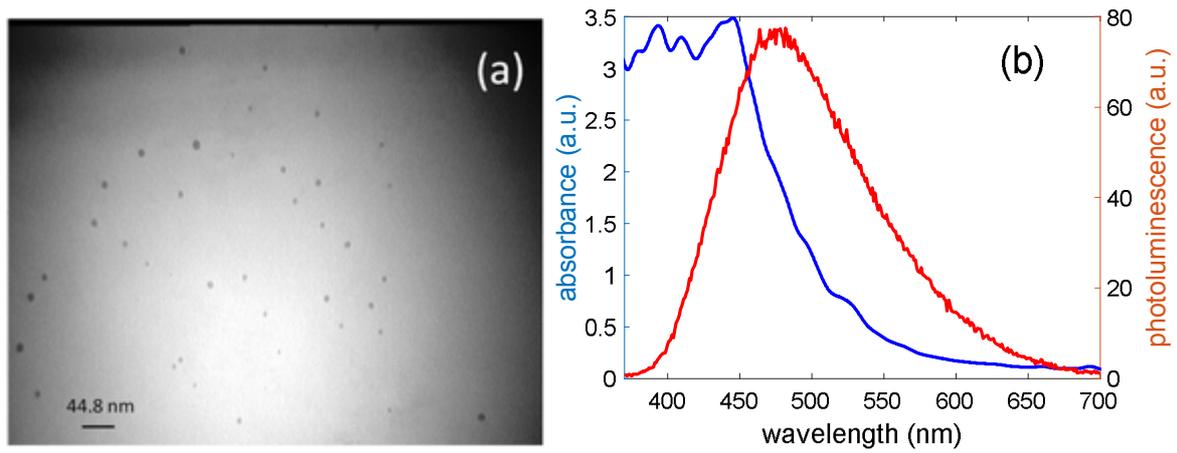

Fig. 1. (a) TEM image, (b) absorbance and PL spectra of the synthesized GQDs.

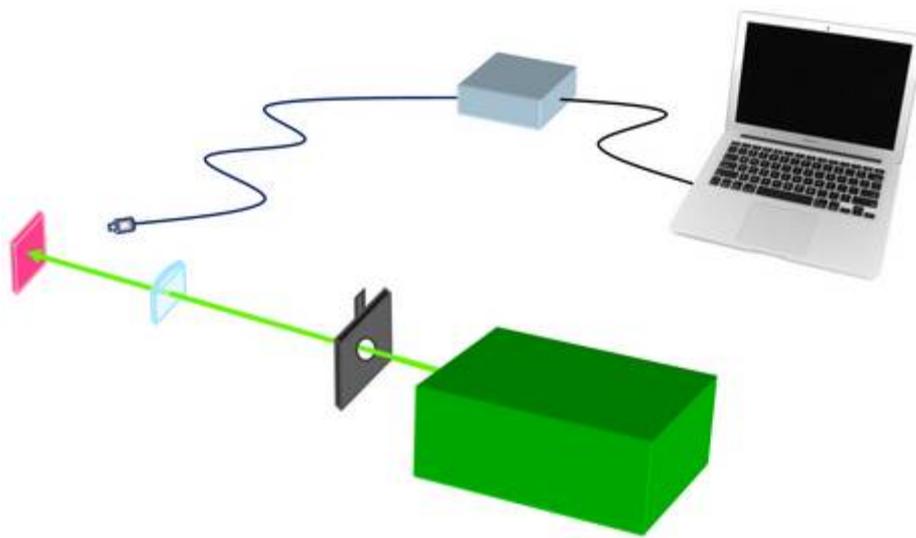

Fig. 2. Sketch of the experimental set up: from down to top are schematics of Nd-YAG laser, aperture diaphragm, cylindrical lens, RL sample, optical fiber, spectrometer and PC.

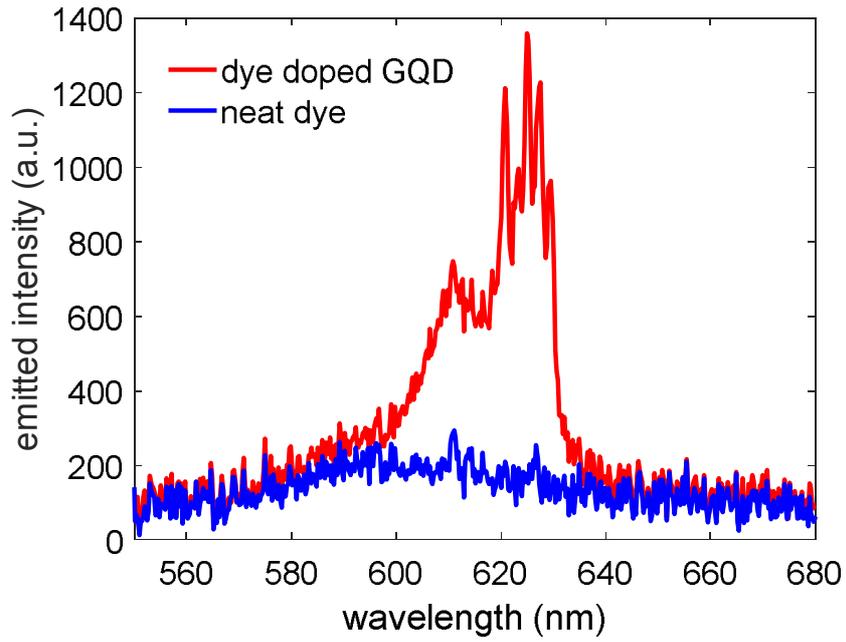

Fig. 3. Emission spectra of sample 1 and that of neat dye at the same pump fluence of 114.84 mJ/cm$^2$.

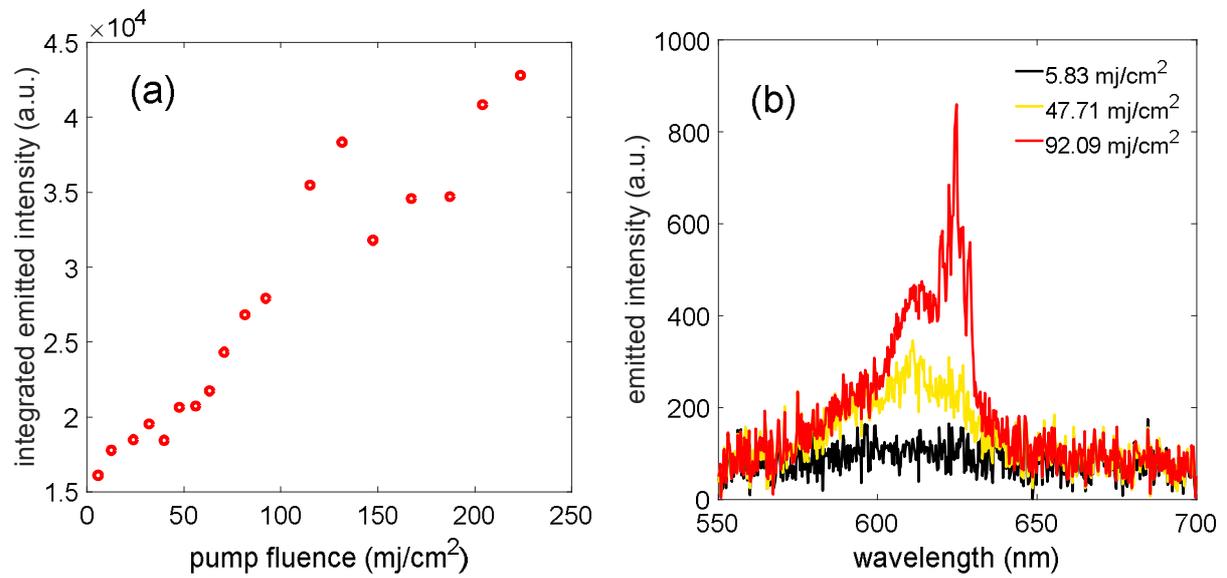

Fig. 4. (a) Plot of spectrally integrated emitted intensity versus pump fluence, and (b) Emission spectra of sample 1 at three different pump fluences.

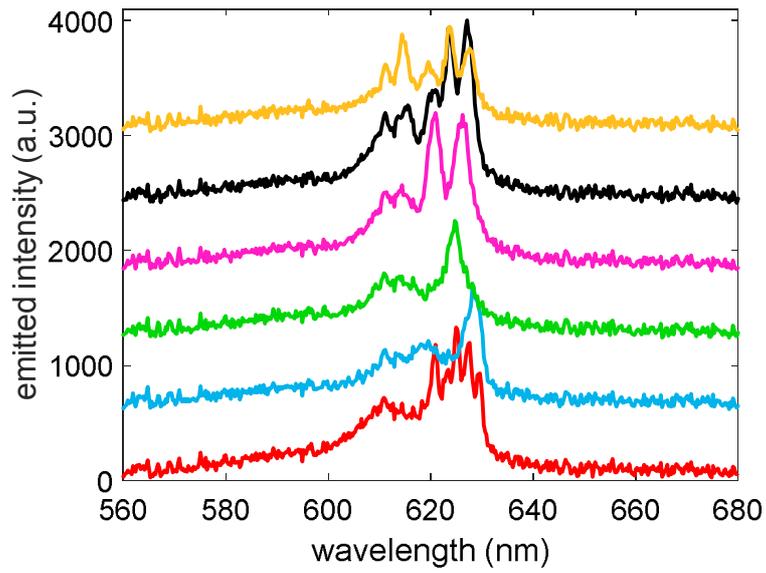

Fig. 5. Emission spectra of sample 1 at six different pump positions.

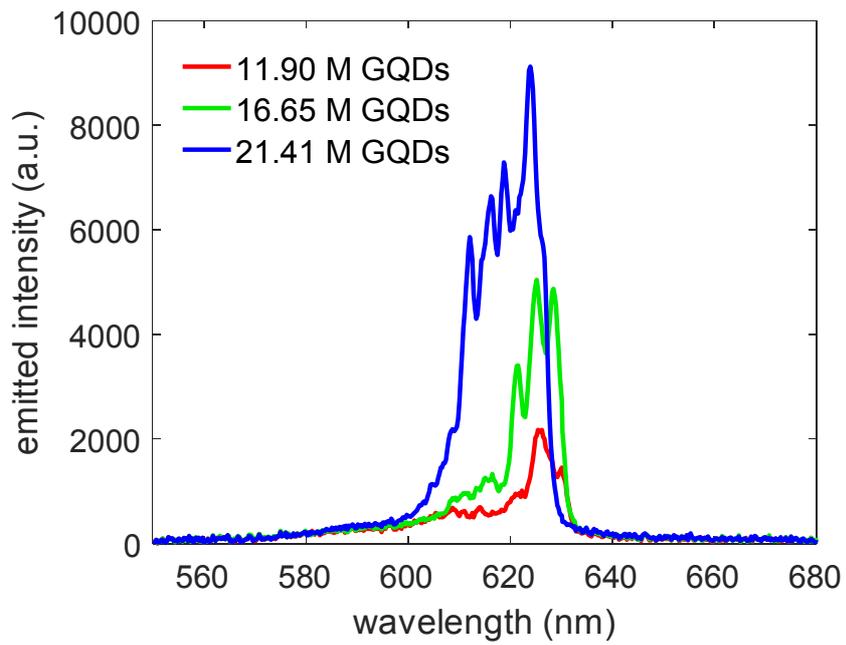

Fig. 6. Emission spectra of samples 1, 2 and 3 at constant pump fluence of 237.17 mJ/cm$^2$.

۹

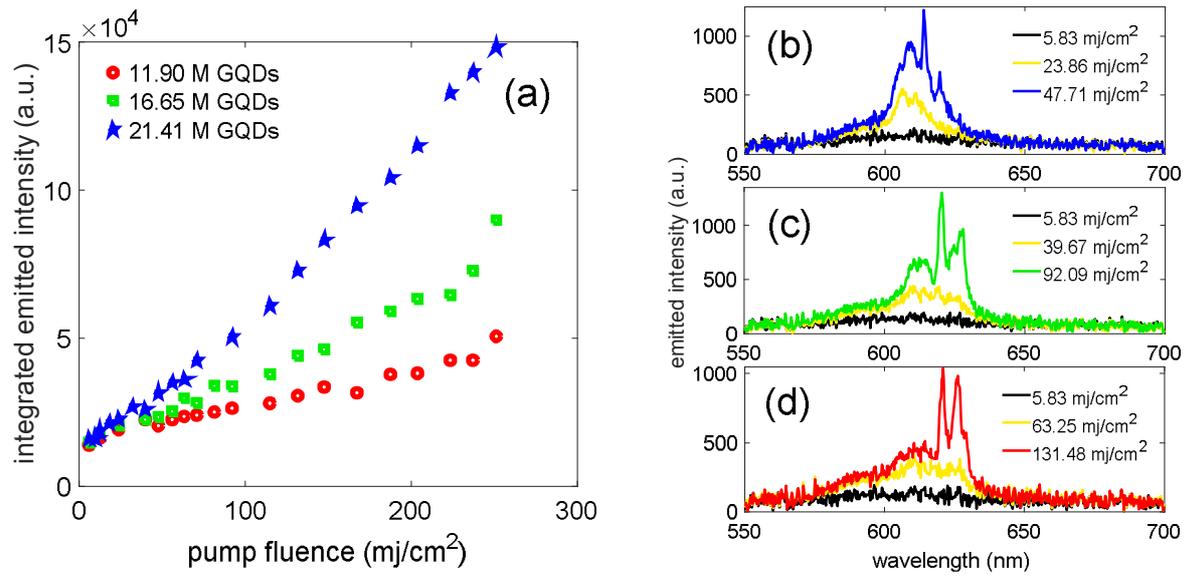

Fig. 7. (a) Plots of spectrally integrated emitted intensity versus pump fluence corresponding to samples 1, 2 and 3. (b), (c) and (d) Emission spectra of samples 3, 2 and 1, respectively at three different pump fluences.